# A study of resistance forces

Gabriel Murariu[1]

[1] *Faculty of Science, University "Dunărea de Jos", Galați, România*

*gabriel_murariu@yahoo.com*

**Abstract**

In this paper we present an example of a simple study of velocity proportional resistance force. Some experimental determinations are presented.

PACS {01.30.Pp. 01.50.My, 01.50.Pa}

## I. Introduction

The first principle of Newtonian mechanics leads to the possibility of experimental description of body interactions.

Before this moment, the space, the time and the interactions are three fundamental notions without any relations between them.

Observing that any interaction-free body moves with constant vectorial velocity, Newton was the first who raised this fundamental observation as a principle level. From that moment, everybody had simple criteria to find out if a body is interaction-free or not, observing his motion.

The most important conclusion was that interactions can be described using the dynamical effects produced and observed.

The second principle introduces a beautiful equation, between the interaction intensity (called force) and the dynamical aspect (given by body acceleration). Newton link these two important aspects of any common interaction as

$$\vec{F}(\vec{r},t) = m\vec{a}(\vec{r},t) \qquad (1)$$

where $\vec{F}(\vec{r},t)$ is the vectorial model of interaction - an experimental description of studied interaction.

The acceleration is defined as

$$\vec{a}(\vec{r},t) = \frac{d^2\vec{r}}{dt^2} = \frac{d\vec{v}}{dt} = \dot{\vec{v}} \qquad (2)$$

and the inertial aspects is measured by the scalar parameter $m$.

In the case of a acting forces system, it has to derive the resultant vector

$$\vec{R}(\vec{r},t) = \vec{F}_1 + \vec{F}_2 + \vec{F}_3 + ... + \vec{F}_n \qquad (3)$$

which can replace the entire system.

Writing the second principle of Newtonian mechanics using the definition (2) of the acceleration, as

$$\vec{F}(\vec{r},t) = m\vec{a}(\vec{r},t) = m\frac{d^2\vec{r}}{dt^2} \qquad (4)$$

leads to idea that the second mechanics principle is expressed as a second rank differential vectorial equation.

In the simple case of a constant force $\vec{F}(\vec{r},t) = \vec{F_0}$, the equation (4) becomes a simple differential equation

$$m\frac{d^2\vec{r}}{dt^2} = \vec{F_0} \quad \Rightarrow \quad \vec{v}(t) = \frac{d\vec{r}(t)}{dt} = \int_{t_0}^{t} \frac{\vec{F_0}}{m} dt + \vec{v_0} = \int_{t_0}^{t} \vec{a} dt + \vec{v_0}$$

or, equivalent

$$\vec{v}(t) = \vec{v_0} + \vec{a}\Delta t \tag{5}$$

where $\vec{v_0}$ is the velocity at the initial moment $t_0$

Further, can be derived the position time dependece using the obtained result

$$\frac{d\vec{r}(t)}{dt} = \vec{v_0} + \vec{a}\Delta t$$

than

$$\vec{r}(t) = \int_{t_0}^{t} (\vec{a}\Delta t + \vec{v_0}) dt = \vec{r_0} + \vec{v_0}\Delta t + \frac{1}{2}\vec{a}\Delta t$$

where $\vec{r_0}$ is the corespondent position at the initial moment $t_0$.

In the simple case on a time constant action force $\vec{F}(\vec{r},t) = \vec{F}(\vec{r})$, then (3) is a differential equation with separable variables.

In a large set of cases, this equation can be solved and can be computed the position time dependence – the motion law

$$\vec{r} = \vec{r}(t) \tag{5}$$

Finally, in our studied case of a velocity dependent force $\vec{F}(\vec{r},t) = \vec{F}(\vec{v} = \dot{\vec{r}})$, then (3) becomes

$$\vec{F}(\vec{v}) = m\frac{d\vec{v}}{dt} \tag{6}$$

This is a first order differential equation for the velocity time dependence

$$\vec{v}(t) = \dot{\vec{r}}(t) \tag{7}$$

This kind of situation is met for a falling in the gravitational field when the air resistance force is present. The resistance force can be consider, in the first approximation level, as proportional with the body's velocity

$$\vec{F_r} = -k\vec{v}$$

The minus sign is given by the opposite orientation of vectors ($\vec{F_r}$ and $\vec{v}$).

In this case, the acceleration can be computed as

$$\vec{a} = \frac{\vec{F_0} - k\vec{v}}{m} \tag{7}$$

where the constant force was noted with $\vec{F_0}$.

In the particular case of a linear motion, the vector arrow can be removed, and further, using the acceleration definition relation, can be derived

$$a = \frac{F_0 - kv}{m} \quad \Rightarrow \quad \frac{dv}{kv - F_0} = -\frac{1}{m}dt \tag{8}$$

The velocity is then given by (considering the initial time moment $t_0 = 0$)

$$v(t) = \frac{F_0}{k} - \xi e^{-\frac{k}{m}t} \qquad (9)$$

where $\xi$ is an integration constant.

If the speed is null at the initial time moment $t_0=0$

$$v_0 = 0$$

the integration constant has to be

$$\xi = \frac{F_0}{k}$$

then, the time dependence of velocity becomes

$$v(t) = \frac{F_0}{k}\left(1 - e^{-\frac{k}{m}t}\right) \qquad (10)$$

This last obtained result leads to finding out the position time dependence, using the velocity definition (7).

$$x = \int_0^t v\, dt = \int_0^t \frac{F_0}{k}\left(1 - e^{-\frac{k}{m}t}\right) dt = x_0 + \frac{F_0}{k}t + \frac{F_0 m}{k^2} e^{-\frac{k}{m}t} \qquad (11)$$

where $x_0$ – is the initial time coordinate.

For a particular case of null velocity initial moment ($v_0=0$), form the referential frame origin ($x_0=0$), position time dependence can be written as

$$x = \int_0^t v\, dt = \frac{F_0}{k}t + \frac{F_0 m}{k^2} e^{-\frac{k}{m}t} = \frac{F_0}{k}\left(t - \frac{m}{k} e^{-\frac{k}{m}t}\right) \qquad (12)$$

This lead to a constant maximum velocity (for $t \to \infty$)

$$v_{max} = \lim_{t \to \infty} \frac{F_0}{k}\left(1 - \frac{m}{kt} e^{-\frac{k}{m}t}\right) \approx \frac{F_0}{k} \qquad (13)$$

Graphical representation for these two functions (time dependence for motion and velocity) are presented in Fig.1 a and b.

**Motion law**

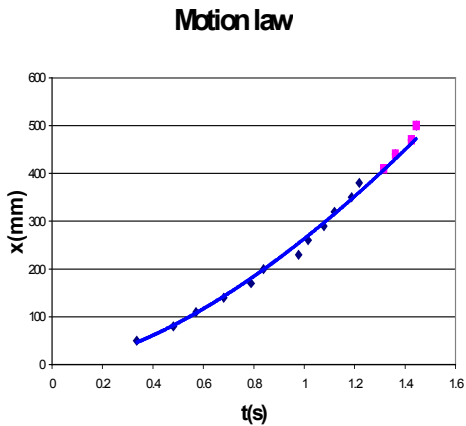

**Velocity law**

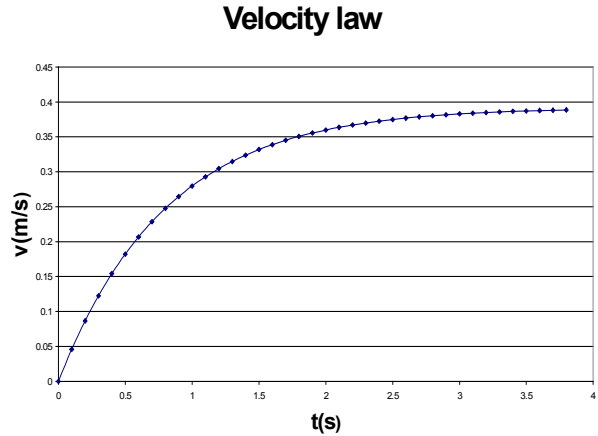

**Fig. 1 a**          **Fig. 1 b**

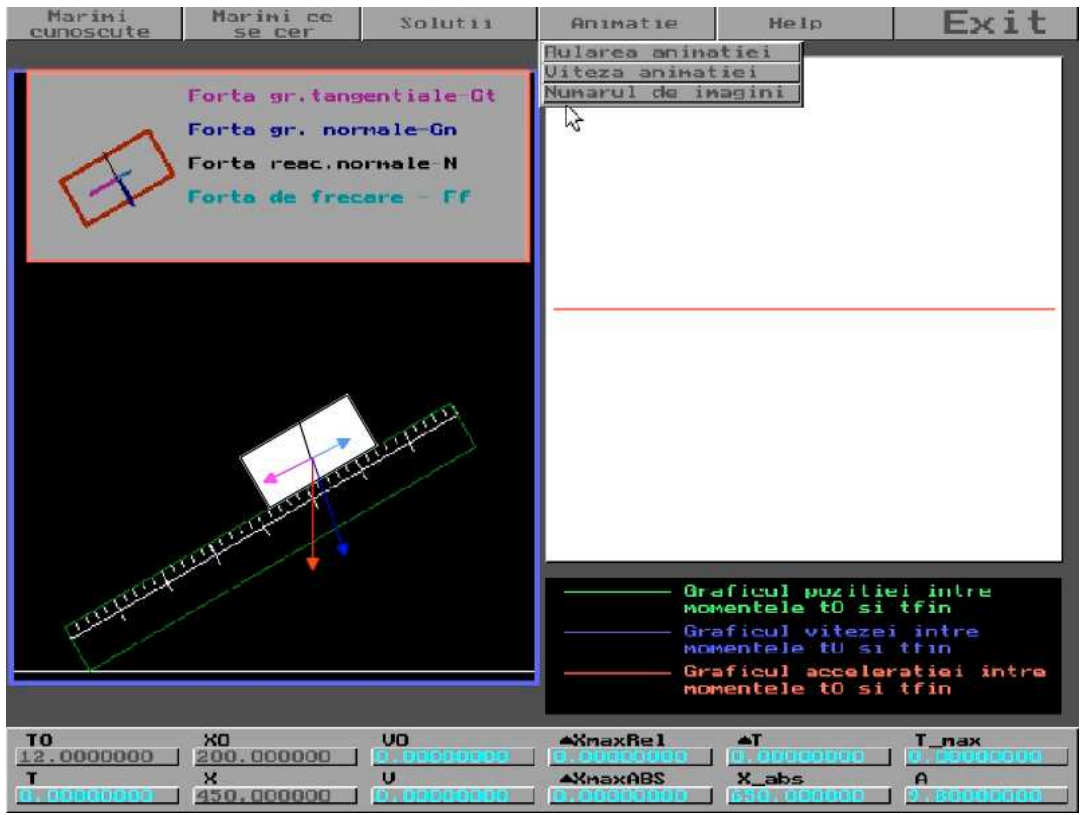

**Fig. 2 – Computer interface**

## II. Experimental device

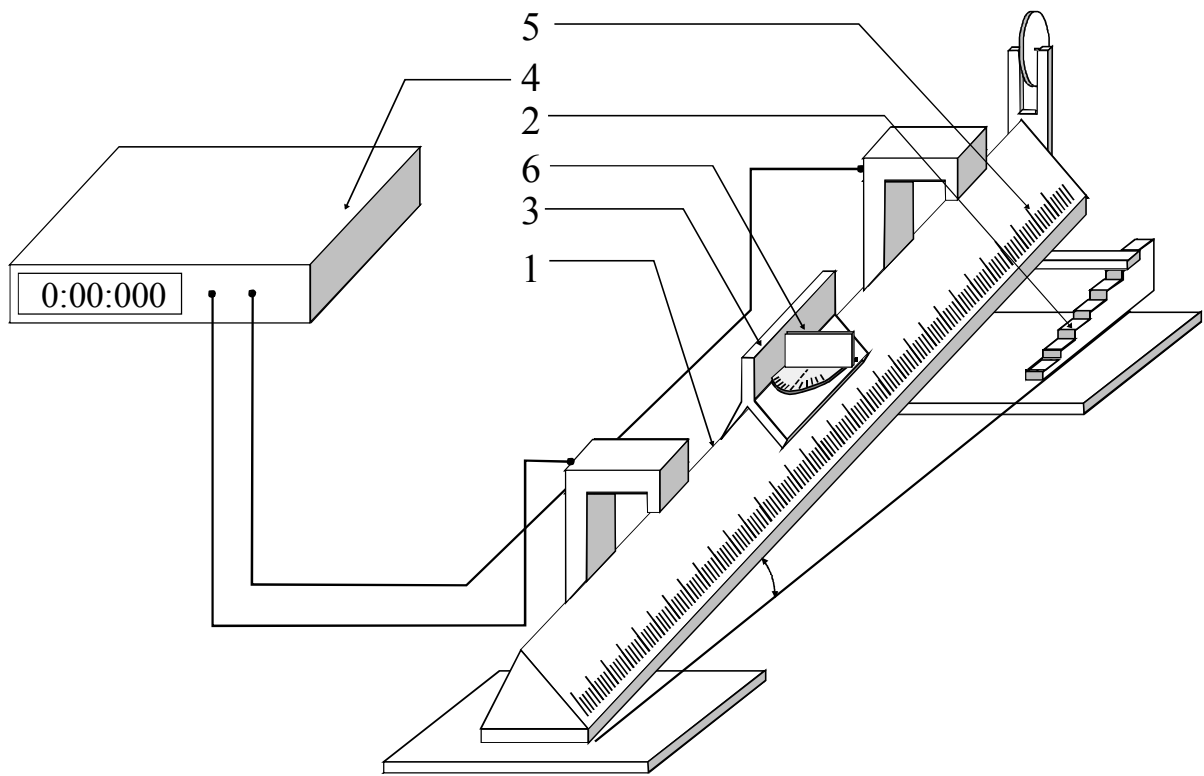

**Fig. 3**

The experimental device is presented further (Fig. 3) and consists from:
- an air track (1)
- lifting device (2)
- mobile body (3)
- electronic chronometer (4) with two optical gates.
- the panel (6) which generates de resistance force can be rotated to different angles.

Varying the lifting angle using (2), can be set the value of tangential weight component. This component is the constant force $F_0$ from

## III. Specific results

Bellow are presented three determinations sets (for three different angles) Representing this dependence $x = x(t)$, can be derived the asymptotic trend and further, using the panta can be computed the resistance force coefficient $k$ as

$$\text{tg}\beta = \frac{F_0}{k} \Leftrightarrow k = \frac{F_0}{\text{tg}\beta} \qquad (14)$$

where
$$F_0 = mg\sin\alpha \qquad (15)$$

The linear asymptothic trend sugests that our hypothesis is corect.

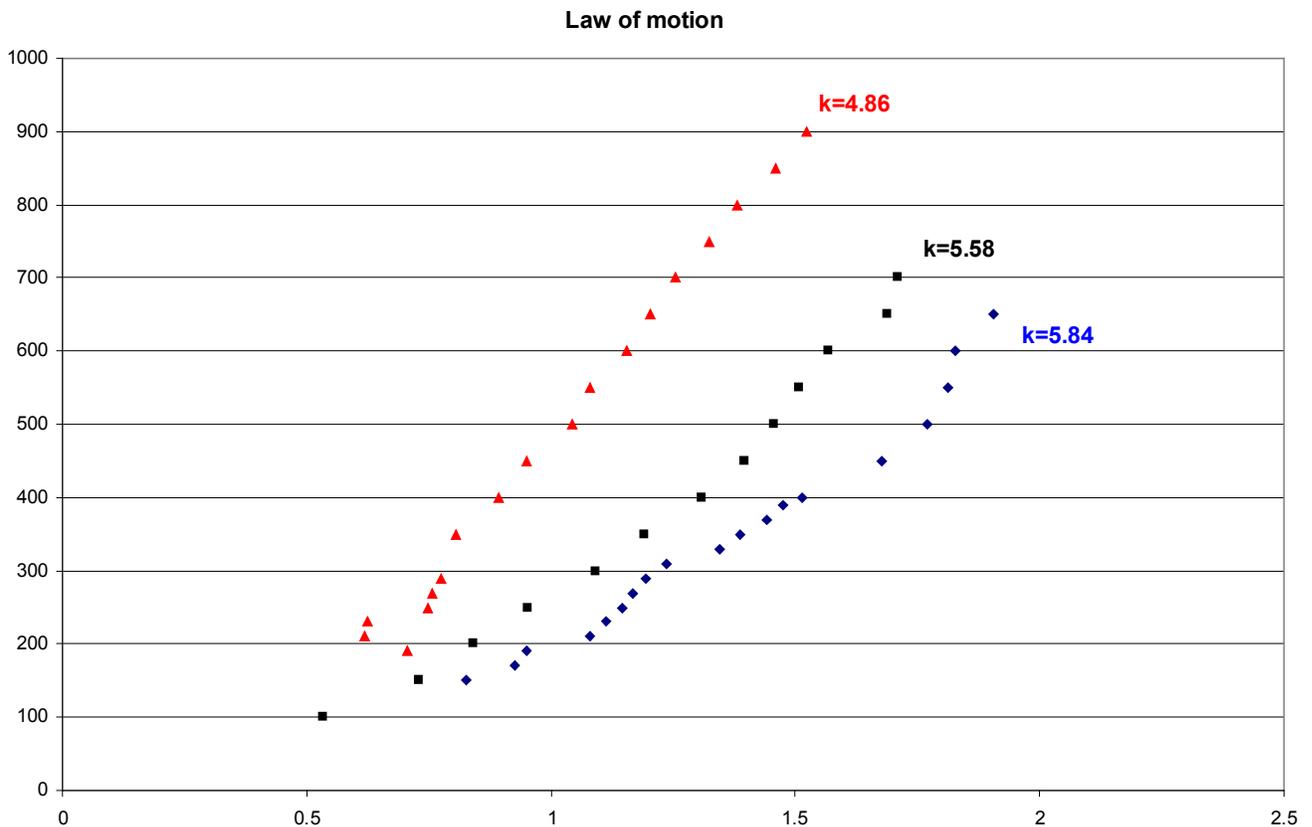

**Fig. 4**

## IV. Acknowledgements

The author wishes to thanks to Sorin Brici for help and fruitful discussions. Special thanks go to Gh. Puşcaşu for his important design work and for his help to succeed in finalizing the asociated computer application.